\documentclass[a4paper,aps,twocolumn,groupedaddress,showkeys,showpacs, 
floats,floatats,floatfix]{revtex4-1}
\usepackage[utf8]{inputenc}
\usepackage{graphicx}  
\usepackage{dcolumn}   
\usepackage{bm}        
\usepackage{natbib}
\usepackage{latexsym}
\usepackage{mathrsfs}
\usepackage{amssymb}
\usepackage{amsmath}
\usepackage{amscd}
\usepackage{color}
\usepackage{verbatim}
\begin{document}
\title{Chaotic and Arnold stripes in weakly chaotic Hamiltonian systems}
\author{M.~S.~Cust\'odio$^1$, C.~Manchein$^2$ and M.~W.~Beims$^1$}
\affiliation{${}^{1}$Departamento de F\'\i sica, 
         Universidade Federal do Paran\'a,
         81531-990 Curitiba, Brazil}
\affiliation{${}^{2}$Departamento de F\'\i sica, Universidade 
        do Estado de Santa Catarina, 89219-710 Joinville, 
        Brazil} 
\date{\today}
\begin{abstract}
The dynamics in weakly chaotic Hamiltonian systems strongly depends on
initial conditions and little can be affirmed about generic
behaviors. Using two distinct Hamiltonian systems, namely one particle
in an open rectangular billiard and four particles globally coupled on
a discrete lattice, we show that in these models the transition from
integrable motion to weak chaos emerges via chaotic stripes as the
nonlinear parameter is increased. The stripes represent intervals  
of initial conditions which generate chaotic trajectories 
and increase with the nonlinear parameter of the system. 
In the billiard case the initial conditions are the injection 
angles. For higher-dimensional systems and small nonlinearities 
the chaotic stripes are the initial condition inside which 
Arnold diffusion occurs.
\end{abstract}
%


\maketitle

In weakly chaotic Hamiltonian systems the domains of regular dynamics
in phase space are very large compared to the domains of chaotic
dynamics \cite{zas-book}. The general characterization of the full
dynamics in such systems is extremely difficult. Starting from
integrable Hamiltonian systems, we show that for weak perturbations
the chaotic trajectories are solely generate inside specific intervals
of initial conditions. These intervals increase with the magnitude 
of the perturbation creating stripes of irregular motion, which are
designed here as chaotic stripes. Chaotic motion for the whole system
is only obtained for larger perturbations {\it via} the overlap of the
chaotic stripes. Results are shown for one particle inside an open
rectangular billiard with rounded corner and four particles coupled
globally on a discrete lattice. Thus, we show that the way weak chaos
is born in the very complex dynamics of the coupled maps, is identical
to what occurs in the simple billiard system. For higher-dimensional
systems the chaotic stripes contain the chaotic channels where Arnold 
diffusion occurs and are designed as Arnold stripes.

\section{Introduction}
\label{Introduction}
The precise description of weak chaos in nonintegrable 
high-dimensional Hamiltonian system is not trivial. When 
starting from an integrable system, the large amount of 
regular structures (invariant tori in two dimensions) 
which remain for small nonlinear perturbations strongly 
influences the dynamics \cite{lichtenberg92}.  
The phase space dynamics is mainly regular with few
chaotic trajectories. In distinction to what happens in 
totally chaotic or ergodic systems, the dynamics is now 
strongly dependent, in a very complex way, on the initial 
conditions (ICs). The present contribution shows a simple 
and clarifying way to analyze the whole dynamics when 
weak chaos is dominant. The dynamics dependence on initial 
conditions is shown in the plot: ICs {\it versus} the 
nonlinear parameter. Chaotic trajectories are 
generated inside intervals of ICs which get larger as the 
nonlinear parameter increases, forming chaotic 
``stripes''-like structures in the plot ICs {\it versus} 
nonlinear parameter. Thus the purpose of the 
present work is to show the way ({\it via} chaotic stripes)
integrable Hamiltonian systems are transformed in weakly 
chaotic systems. Since chaotic stripes are expected 
to exist for a large class of Hamiltonian systems, 
results are shown here for two entirely different systems. 

The first system is the two 
dimensional open rectangular billiard where
the corners have been modified, they are rounded.
Such open billiard, but with straight corners and a rounded 
open channel, was studied recently \cite{marcelo11-1}. It 
was shown that the rounded open channel generates a 
chaotic dynamics inside the stripes and thus inside the 
billiard. The chaotic motion is characterized based on 
exponential decays of escape times (ETs) statistics 
\cite{altmann06} inside the stripes.  
We would like to mention some examples of two-dimensional 
billiards where boundaries have been modified: edge 
roughness in quantum dots \cite{libisch09}, soft billiards 
\cite{hercules11-1,baldwin88,knauf89,rom-kedar03}, edge 
corrections \cite{blumel05} in a resonator, effects of 
soft walls \cite{hercules08}, rounded edges 
\cite{marcelo10,wiersig03}, deformation of dielectric 
cavities \cite{bogo08}, location of the hole \cite{buni10}, 
among others.

The second example considered here comes from the nonlinear 
dynamics, namely $4$ particles which are globally coupled 
on a discrete lattice. This system was 
extensively studied in \cite{kk90,kk92,kaneko94}, and the 
dynamics characterized by the existence of an ordering 
process called clustering. In the context of the present 
work, this model serves to show that chaotic stripes exist 
in a higher-dimensional system and are related to the 
Arnold web, 
or stochastic web \cite{zas-book}. Different from the 
billiard case, the chaotic motion will be characterized 
by the Finite Time Lyapunov 
Exponent (FTLE). Starting from a chaotic trajectory, 
regular structures induce sticky motion 
\cite{dettmann11,contopoulos10,zas02}, 
the convergence of the FTLEs is affected and the dynamics 
becomes strongly dependent on ICs. Recent methods
\cite{steven07,cesar1,hercules08,cesarPRE09,manchein09,cesar2},
which use higher-order cummulants of Gaussian-like 
distributions of the FTLEs to characterize the whole 
dynamics via sticky motion, do not work quite well 
for very small nonlinear perturbations since the 
distributions may be multimodal \cite{lopes05}.

Although we are considering two totally 
distinct physical systems, the common feature is that 
the appearance of chaotic trajectories for weak 
perturbations occurs {\it via} the same process,
the chaotic stripes. In other words we 
are interested to understand how the transition from 
integrable motion to weakly chaotic regime happens as the 
parameter of nonlinearity is increased. This occurs for the very simple
billiard system and for the complex coupled maps,
for which chaotic stripes are shown to be related
to the Arnold web. Since the second model contains 
coupled standard maps, results should be valid for 
a large class of dynamical systems.

The paper is organized as follows. Section~\ref{model} presents 
the model of the open billiard system and shows numerical results 
for the ETs decays and the existence of stripes in emission angles. 
Section~\ref{maps} presents the coupled 
map lattice model, shows the existence of the stripes and their 
relation to the Arnold web. In Section~\ref{conclusions}
we present our final remarks.

\section{Stripes in Open Billiards}
\label{model}

\begin{figure}[htb]
 \unitlength 1mm
 \begin{center}
 \includegraphics*[width=8.5cm,angle=0]{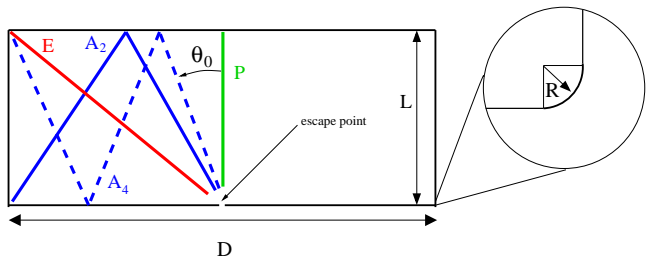}
 \end{center}
\caption{(Color online) The rectangular billiard, with dimension
$L\times D$, has four rounded corners (with radius $R$, see
magnification of just one border). The escape point with aperture 
$a$ lies exactly in the middle of the billiard. Initial angle 
$\theta_0$ and, schematically, the shortest escape trajectories 
($P,E,A_2$ and $A_4$) are shown. In all 
simulations we use $L=4$ and $D=10$.}
  \label{Model-Corner}
  \end{figure}

The model used is shown in Fig.~\ref{Model-Corner}, which 
is a rectangular billiard with dimension $L\times D$. The
four corners are assumed to be rounded with radius $R$.
{ For $R/L=0$ some orbits ($P,E,A_2,A_4$) are shown 
schematically in Fig.~\ref{Model-Corner}, whose inital angles are
$\theta_0^{(n)}=\arctan{\left[\frac{D}{2nL}\right]}$.  
For the closed billiard these are periodic orbits with period 
$q=2^n$ ($n=1,2,3,\ldots$).

The closed rectangular billiard is integrable in the absence of 
corner effects ($R=0$). All Lyapunov exponents are zero and the 
dynamics can be described in terms of invariant tori. Such 
regular dynamics is shown in the the phase-space  of 
Fig.~\ref{Phase-Space}(a), where the collision angle $\theta$ is 
plotted as a function of horizontal coordinate $x$.  Tori with 
irrational winding numbers are the straight lines parallel to 
the $x$-axis. Tori with rational winding numbers are periodic 
orbits and are the marginally unstable periodic orbits (MUPOs) 
from this problem. When opening up the billiard, ICs 
which start on an irrational torus will certainly escape the 
billiard after some time, but trajectories which start exactly on 
a rational tori will never leave the billiard, except for 
those ICs which match the opening channel. 
 \begin{figure}[htb]
 \unitlength 1mm
 \begin{center}
 \includegraphics*[width=8.5cm,angle=0]{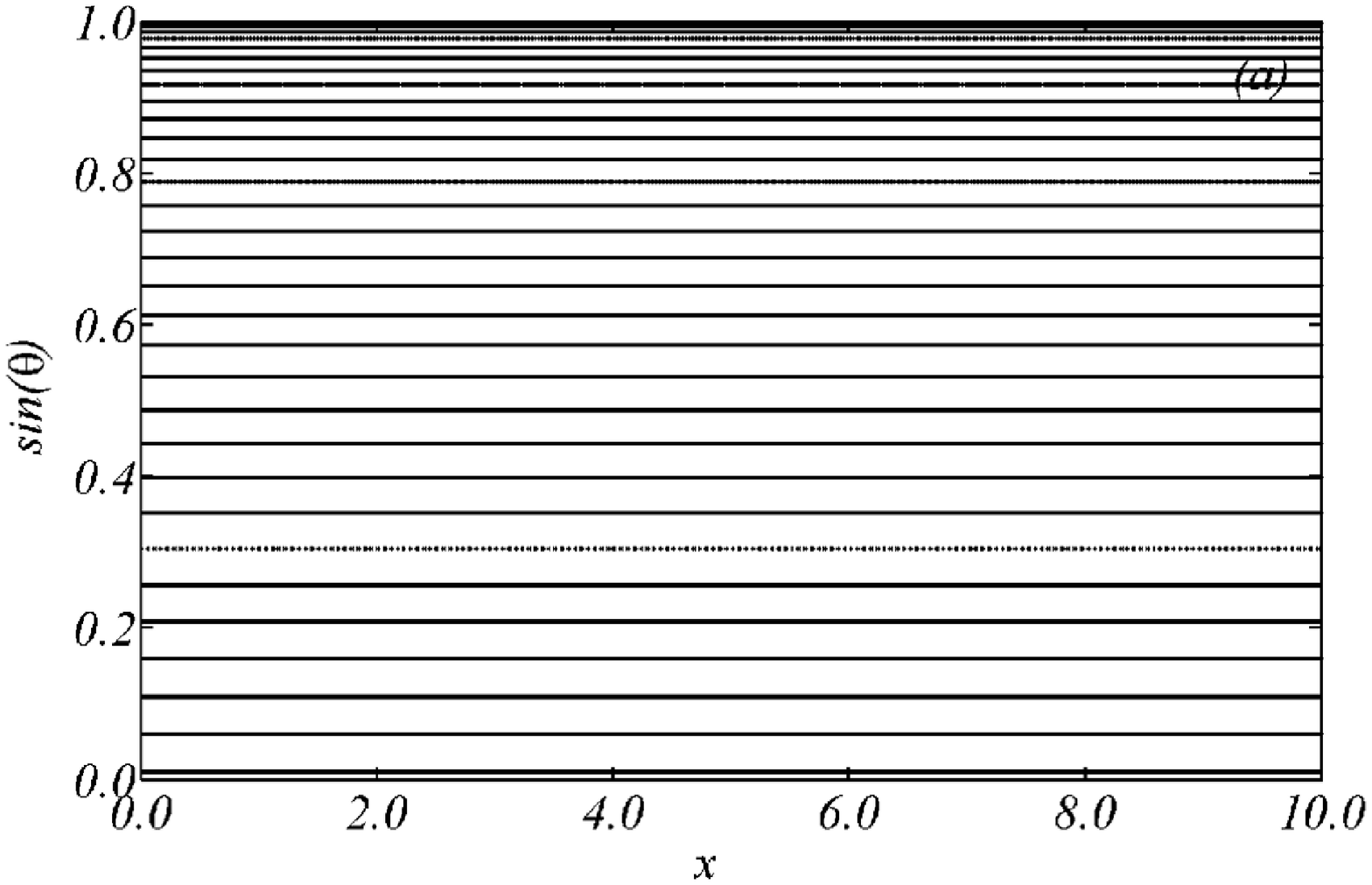}
 \includegraphics*[width=8.5cm,angle=0]{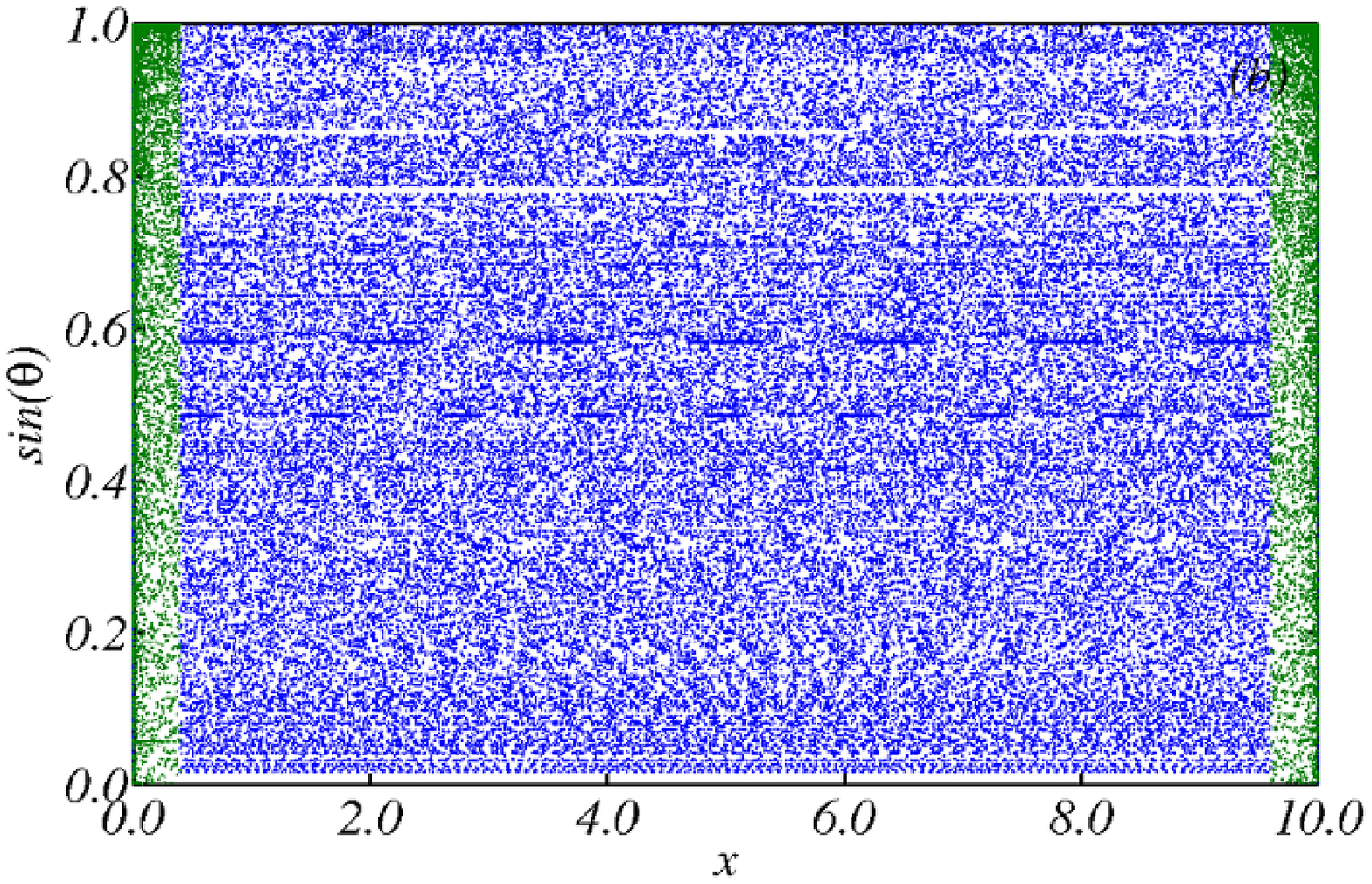}
 \end{center}
\caption{(Color online) Phase-space dynamics for (a) $R=0$ and 
(b) $R/L=0.1$ (one IC): Green (light gray) points 
occur when the trajectory collides with the rounded corner and 
defines the scattering region (SR) or chaotic region. Blue (dark 
gray) points occur when the trajectory collides with the vertical 
and horizontal parallel walls.}
  \label{Phase-Space}
  \end{figure}
Figure \ref{Phase-Space}(b) shows the phase-space dynamics when
$R/L=0.1$ starting from one IC $\theta_0=0.86$. Blue 
points are related to points along the trajectory which collide
with vertical and horizontal parallel straight walls, while 
green points are 
those which collide with the rounded corner. What happens is that 
the trajectory starts on an irrational torus, travels along 
this torus (blue horizontal lines) until it collides with one of 
the rounded corners (green points) and is scattered there to a
new torus. It travels on this new tori until it is scattered 
again to another tori. This procedure repeats itself until 
the whole phase space is filled. This is exactly what happens 
in Fig.~\ref{Phase-Space}(b), obtained using just one initial 
condition. The rounded corner is called scattering region (SR) 
and generates the chaotic dynamics. Although there are no 
regular islands in phase space, sticky motion occurs due to 
the existence of MUPOS, as explained in details in 
\cite{marcelo11-1}. For some of the horizontal 
lines in Fig.\ref{Phase-Space}(b), there are some regions 
without points. These are trajectories very close to the
periodic orbits (closed billiard case) shown in 
Fig.~\ref{Model-Corner}, and before they can visit all the 
points on the horizontal line, they are scattered by the SR
to another horizontal line.}

In order to analyze the effect of rounded corners on the emission
angles dynamics we start simulations at times $t=0$ 
from the open channel, uniformly distributed from 
$x_0=D/2$, with an initial angle $\theta_0$ towards 
the inner part of the billiard with velocity $|\vec v|=1$. 
Elastic collisions are assumed at the billiard boundaries and at 
the rounded corners of the open channel. For each of the $10^5$ 
IC  distributed uniformly in the interval 
$1.00\times 10^{-2}\lesssim\theta_0\lesssim 1.57$ we wait until the particle 
leaves the billiard and record $\theta_f$. In all cases $a=0.4$.
 
\begin{figure}[htb]
 \unitlength 1mm
 \begin{center}
\includegraphics*[width=8.5cm,angle=0]{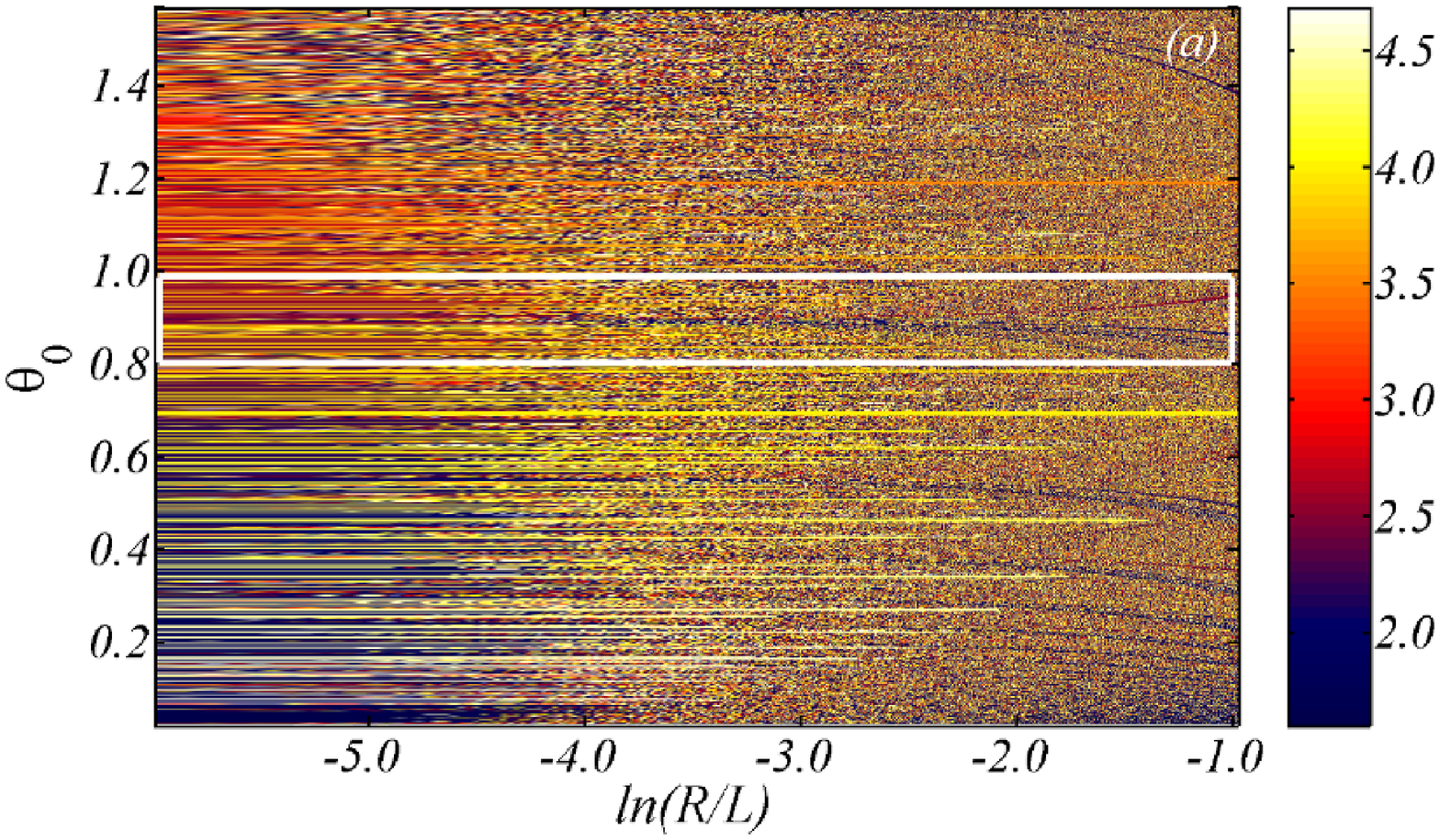}
\includegraphics*[width=8.5cm,angle=0]{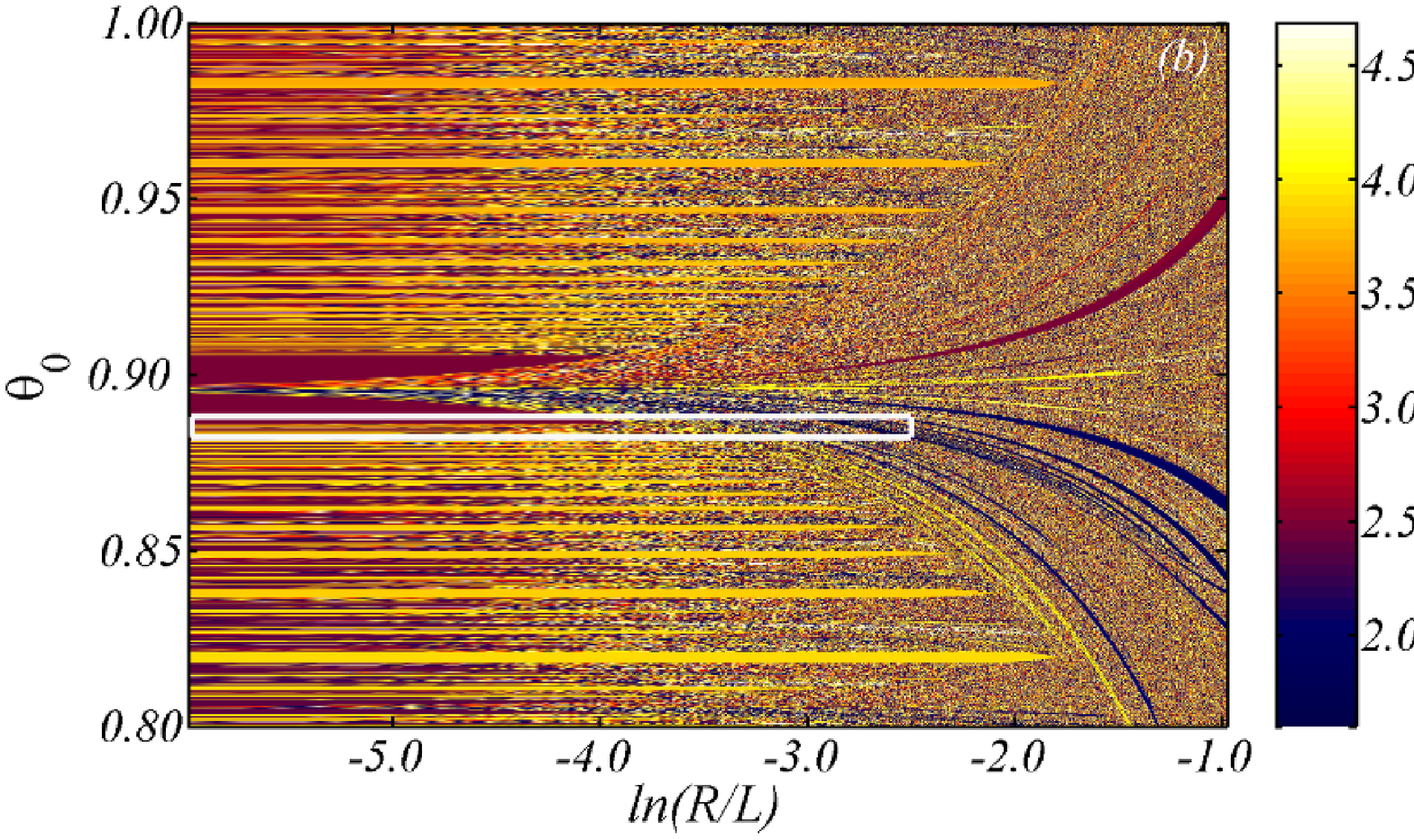}
\includegraphics*[width=8.5cm,angle=0]{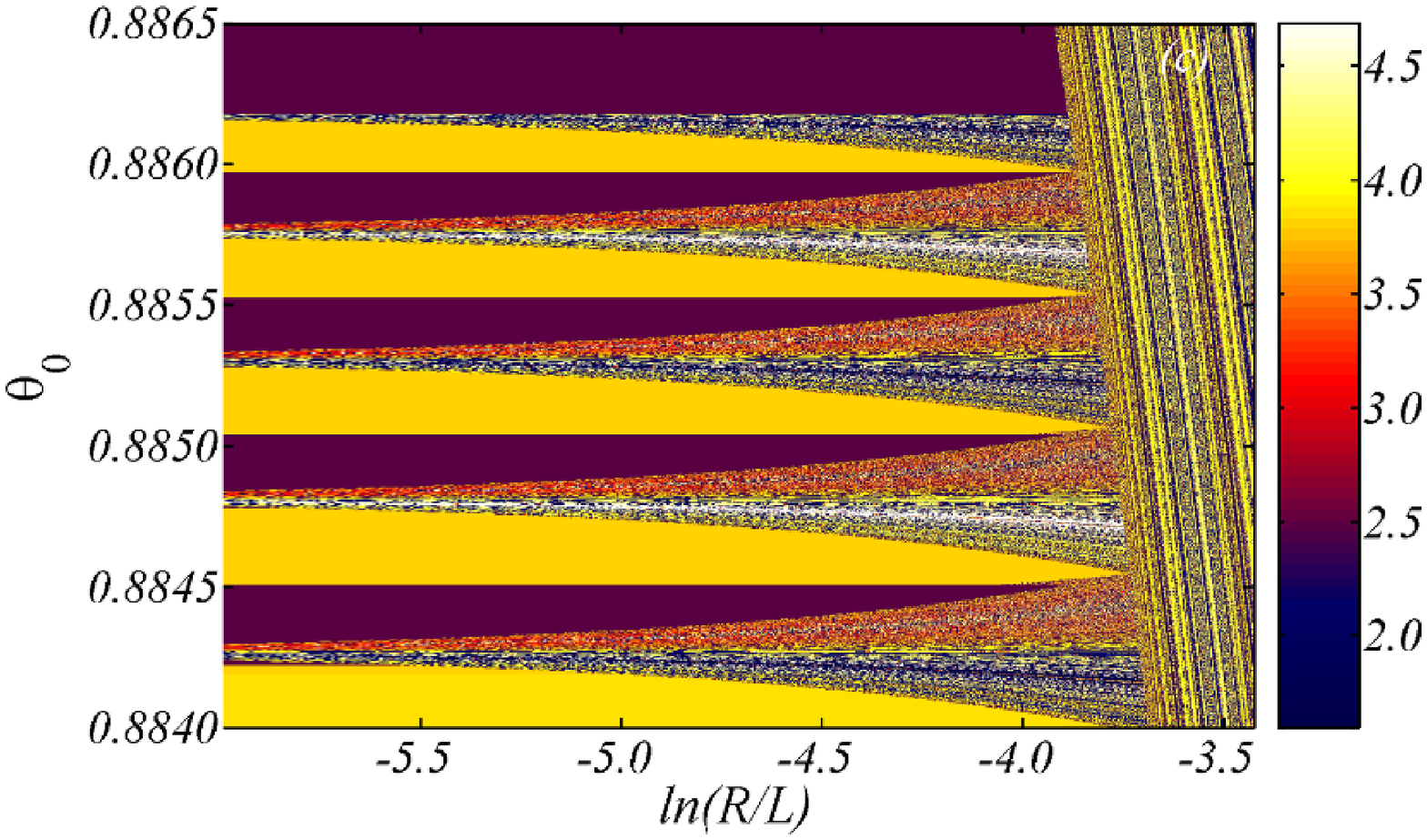}
 \end{center}
\caption{(Color online) (a) Escape angle as a function of 
$\log{(R/L)}$ and $\theta_0$, (b) and (c) are magnifications.}
  \label{time}
  \end{figure}

The complex dynamics generated by the rounded corners becomes 
apparent when the escape angles $\theta_f$ are plotted as 
function of the initial incoming angle $\theta_0$ and for different 
ratios $R/L$. This is shown in Fig.~\ref{time} which was obtained 
by using $10^3\times 10^3$ points in the interval
$2.0 \times 10^{-3}\le R/L\le 0.35$ [$-6.0\le \log{(R/L)}\le -1.0$]. 
Figure \ref{time}(a) shows $\theta_f$ as a function of $\log{(R/L)}$ 
and $\theta_0$. Each color is related to one emission angle $\theta_f$ 
(see the colorbar on the right: dark blue $\to$ red $\to$ yellow $\to$ 
white in the colorscale and dark gray to white in the grayscale). These 
emission angles vary between $\theta_f\sim 1.4$ (almost horizontally to 
the left) and  $\theta_f\sim 4.5$ (almost horizontally to the right). 
It is possible to observe [better seen in the 
magnifications, Figs.~\ref{time}(b) and (c)] that there are always 
intervals of ICs which leave to the same $\theta_f$ (same color). As 
$R/L$ increases, they appear as horizontal lines with the same color,
and are called here as ``isoemissions stripes''. As $R/L$ increases 
more and more, 
some isoemissions stripes survive while others are destroyed or mixed. 
Examples of such stripes are seen in Fig.~\ref{time}(a) 
around $\theta_0\sim 0.7$ (yellow), in 
Fig.~\ref{time}(b) around  $\theta_0\sim 0.9$ (brown), among 
many others. In general, the emission angles show a very rich 
dynamics due to the increasing rounded corners, alternating 
between all possible colors. 

{Figure \ref{time}(b) shows a magnification from 
Fig.~\ref{time}(a). The magnification is taken around the
isoemissions stripe related to the trajectory $E$ 
($\theta_0\sim0.89$) shown in Fig.~\ref{Model-Corner}. We 
observe that a new stripe emerges symmetrically from 
the middle of the isoemissions stripe.  
Different from the isoemissions stripe, inside this new
stripe inumerous distinct escape angles (colors) appear.
This new stripe is called ``chaotic stripe'' for reasons 
which become clear later.
The width of the chaotic stripe increases linearly 
with $R/L$. Outside the isoemissions stripe a large amount 
of smaller stripes with distinct angles can be observed. 
In order to show this in more details and to explain the 
physics involved in the emission angles, we discuss next a 
magnification of Fig.~\ref{time}(b) (see white box).
This is shown in Fig.~\ref{time}(c), where a sequence of 
isoemissions (yellow and purple) and chaotic (all colors) 
stripes appear. 
ICs which start inside the chaotic stripes collide, at 
least once, with the rounded corner so that distinct emission
angles can be observed. Different from what is observed from 
border effects \cite{marcelo11-1}, here the sequence of 
isoemissions stripes in Fig.~\ref{time}(b), and the 
corresponding multicolor 
chaotic stripes from Fig.~\ref{time}(c), are {\it not} born 
at the boundary between the purple (black) and orange (white) 
escape angles at $R/L\sim 0$. The rich and complex dynamics is 
always generated from ICs starting inside the 
chaotic stripes, so that a large amount of different colors 
in the emission angles appears, showing that tiny changes or 
errors in the initial angle may drastically change the emission 
angle. The location of the chaotic stripes itself is not 
self-similar, but {\it inside} the chaotic stripes the 
self-similar structure is evident. Chaotic stripes which 
emerge at different initial angles at $R/L\sim0.0$ increase 
their width with $R/L$ and start to overlap for $R/L\sim0.05$, 
where the dynamics of strongly sensitive to initial angles 
$\theta_0$. The width of the stripes increase with $R/L$ and 
allows us to characterize the different domains of the dynamics.

The dynamics generated by starting with IC {\it inside} the 
chaotic stripes is the consequence of trajectories which 
collide with the rounded corner and the chaotic motion 
appears. The collision with the rounded border
was checked numerically (not shown) and the chaotic motion
has to be demonstrated. To do 
this we use the ETs statistic defined \cite{altmann05} by
$Q(\tau) = \lim_{N\to\infty}\frac{N_{\tau}}{N},$ where $N$ is the 
total number of recurrences and $N_{\tau}$ is the number of 
recurrences with time $T\ge\tau$. The time $T$ is recorded 
when the trajectory returns to the recurrence region, which 
is the open channel. In our model, when a 
trajectory returns to the open channel it leaves the 
billiard. Thus, since we have an open system, the number 
of recurrences is counted over distinct ICs. In chaotic 
hyperbolic systems the ETs 
statistic decays exponentially while systems with stickiness 
it decays as a power law $Q(\tau)\propto \tau^{-\gamma_{esc}}$ 
with $\gamma_{esc}>1.0$ being the scaling exponent. Usually it 
is assumed to have stickiness when a power-law decay with 
$\gamma_{esc}>1.0$ is observed for two decades in time. 
Power-law decays with $\gamma_{esc}=1.0$ \cite{bauer90,marcelo11-1}
appear in integrable systems, which is the case of our model 
when $R/L=0.0$.
 \begin{figure}[htb]
 \unitlength 1mm
 \begin{center}
\includegraphics*[width=8.7cm,angle=0]{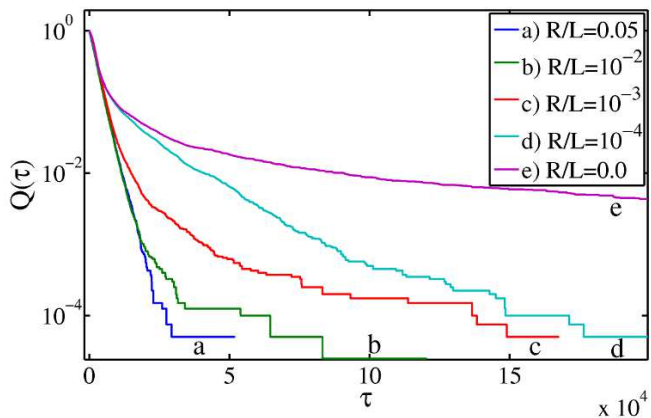}
 \end{center}
\caption{(Color online) Semi-log plot of $Q(\tau)$ for distinct 
values of $R/L$. Each curve was obtained using $2.0 \times 10^4$ ICs.}
  \label{QtR}
  \end{figure}
Figure \ref{QtR} displays in a semi-log plot the quantity $Q(\tau)$ 
obtained using different values of the ratio 
$R/L=0.0,1.0\times 10^{-4},1.0\times 10^{-3},1.0\times 10^{-2},5.0\times 10^{-2}$,
where $L$ is kept fixed. ICs are $1.00 \times 10^{-5} \le \theta_0\le1.57$ and
$-1.00 \times 10^{-5} \le \theta_0\le -1.57$. Straight lines in this plot are
exponential decays. For the integrable case $R/L=0.0$ (curve $e$) a 
power law decays occurs with $\gamma_{esc}\sim 1.0$. Trajectories close 
to the periodic orbits inside the rectangle generate the power-law 
decay with $\gamma_{esc}=1.0$. Small corner effects, 
$R/L=1.0 \times 10^{-4}\to 1.0 \times 10^{-3}$ (see curves $d\to c$), change the qualitative 
behavior of $Q(\tau)$ so that regular trajectories close to MUPOs 
generate, for small period of times, a power-law decay with 
$1.0<\gamma_{esc}<2.0$. For a detailed discussion of a similar behavior 
we refer the reader to \cite{marcelo11-1}. However, for 
$R/L= 1.0 \times 10^{-2} [\log(R/L)=-4.6], 5.0 \times 10^{-2} [\log(R/L)=-3.0]$ 
(curves $b,a$), which correspond to values from Fig.~\ref{time}(c), 
a straight line decay is observed, characterizing the transition 
to an almost totally chaotic motion when stripes overlap.

Summarizing results for the billiard model, we 
showed that specific intervals of ICs, called chaotic stripes,
which increase linearly with the nonlinear parameter, generate 
chaotic trajectories. Outside these stripes we have the isoemissions
stripes, for which the ICs generate a regular motion. For smaller 
values of the parameter, we have a mixture of regular and chaotic 
motion, i.e.~a mixture of a large amount of isoemissions stripes 
and few chaotic stripes, and the system is weakly chaotic. Only when 
the nonlinear parameter increases, so that all chaotic stripes overlap, 
the whole system becomes chaotic.

\section{Stripes in Coupled maps}
\label{maps}
At next we analyze the IC dependence of 
the dynamics of $4$ particles globally coupled on a 
discrete lattice. It has a $8$-dimensional phase space
and particles are coupled on a unit circle, where the state 
of each particle is defined by its position $2\pi x^{(i)}$ 
and its conjugate momentum  $p^{(i)}$.
The dynamics on the lattice obeys \cite{kk90}
\begin{eqnarray}
    \left\{
    \begin{array}{ll}
        p_{t+1}^{(i)} = p_t^{(i)} 
     +\frac{K}{2\pi\sqrt{3}}\sum_{j=1,j\not=i}^{N=4}
    \sin[2\pi(x_t^{(j)}-x_t^{(i)})], \\
\\
	x_{t+1}^{(i)} = x_t^{(i)} + p_{t+1}^{(i)}, 
    \end{array}
    \right.
    \label{gc}
\end{eqnarray}
where mod $1$ must be taken in the variables $p^{(i)},x^{(i)}$, 
and $i=1,\ldots,N=4$. For $K>0$ the interaction between two 
particles $i$ and $j$ is attractive \cite{kk92}. For this 
coupling the total momentum $P_T=\sum_{j=1}^{N=4} p^{j}_t$ is 
preserved. This model has $8$ Lyapunov exponents
which come in pairs. The reasons for using such a model
are: (i) these are coupled standard maps, thus their 
properties should be valid for a large class of physical 
systems; (ii) the phase space dynamics is $8$, thus it 
can be used to show that, in case results are similar to 
those from Section \ref{model}, they are valid for higher 
dimensions and connect the chaotic stripes to the Arnold 
web.

An interesting way to describe the complicated dynamics 
dependence on ICs is shown in the plot $K\times p_0^{(2)}$ 
from Fig.~\ref{coupled}. In colors is plotted the largest 
FTLE after $10^4$ iterations. Initial conditions are chosen 
on the invariant structure $P_T=0.0$ and $X_{CM}=0.0$. This 
plot shows, as a function of $K$, those ICs $p_0^{(2)}$ which 
generate the regular, mixed or chaotic motion. While black 
colors describe initial conditions with generate a regular 
motion, with zero FTLEs, cyan, red to yellow colors are 
obtained from ICs which generate increasing positive FTLEs.
 \begin{figure}[htb]
 \unitlength 1mm
 \begin{center}
\includegraphics*[width=8.7cm,angle=0]{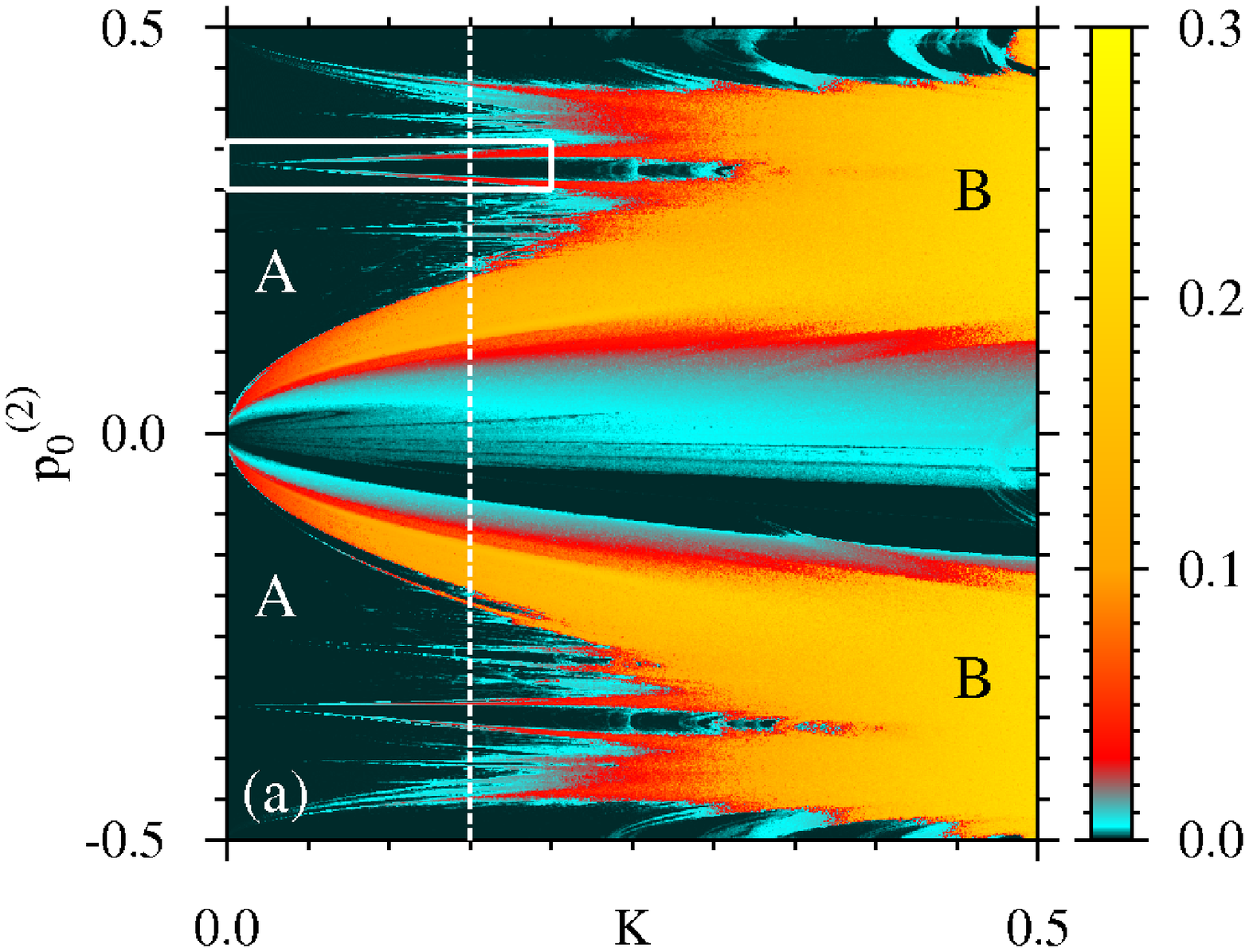}
\includegraphics*[width=8.7cm,angle=0]{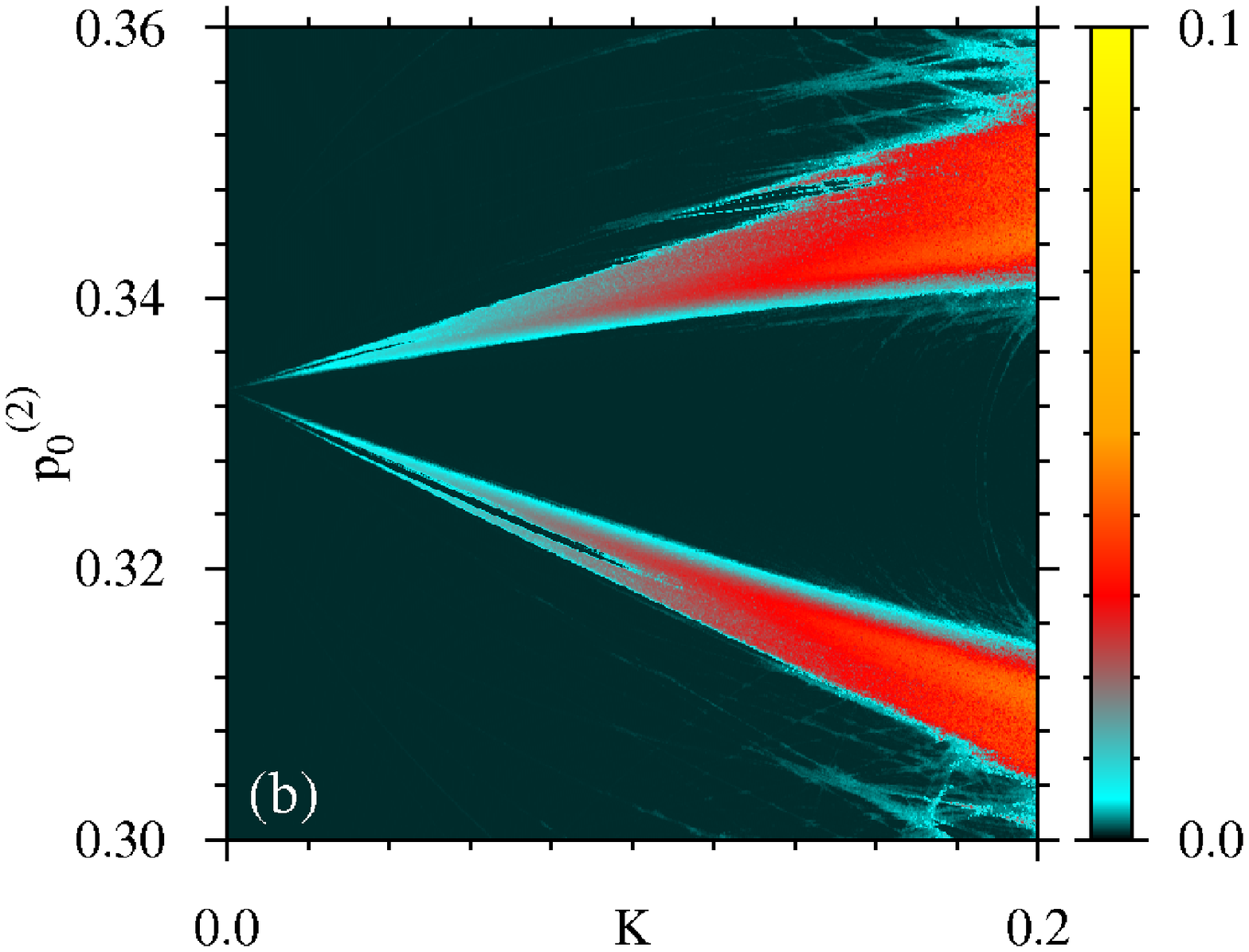}
 \end{center}
\caption{(Color online) Grid of $1000\times 1000$ points 
with the FTLEs as a function of $K\times p_0^{(2)}$.} 
  \label{coupled}
  \end{figure}
A rich variety of structures is observed. Essentially two distinct larger
regions of motions are observed and demarked in the plot as: (A) where 
FTLEs are smaller and (B) where FTLEs are larger. In between the FTLEs mix 
themselves along complicated and apparently fractal structures. For $K\to0$ 
FTLEs also go to zero inside region (A). However there are some stripes, 
which emanate from $K\sim 0$ for which FTLEs are larger. Two larger 
stripes are born close to $p_0^{(2)}\sim 0.0$ and growth symmetrically 
around this point as $K$ increases.  ICs which start inside
the stripes generate chaotic trajectories. Thus these stripes correspond
to the chaotic stripes found in the billiard case in Sec.~\ref{model}.
Along the line $p_0^{(2)}\sim 0.0$ 
the dynamics is almost regular for $K<1$, becoming chaotic for larger 
values of $K$. This picture shows us clearly that for the same $K$ 
value, different ICs (in this case $p_0^{(2)}$) have 
distinct FTLEs. Zero FTLEs are related to regular trajectories,
larger FTLEs are related to chaotic trajectories and intermediate
FTLEs are related to chaotic trajectories which touched, for 
a finite time, the regular structures from the high-dimensional 
phase space. Such regular structures can be global invariants 
(as $P_T$) or local, or collective ordered states which live in 
the high-dimensional phase space. As trajectories itinerate 
between ordered and random states, the regular structures affect 
locally the FTLEs inducing sticky motion, so that the 
corresponding FTLEs decrease. Thus each point in the mixed plot 
from Fig.~\ref{coupled}(a) with and intermediate FTLE for a 
fixed $K$ value, is necessarily related to a trajectory which 
suffers sticky motion. Figure \ref{coupled}(b) is a 
magnification of Fig.~\ref{coupled}(a) and shows similar stripes 
structures observed in the billiard case. As the nonlinear 
parameter $K$ increases, the width of the chaotic stripes 
increase and they start to overlap, generating the totally 
chaotic motion, but only for higher $K$ values (not shown).

It is very interesting to observed that the chaotic stripes 
define the channels of chaotic trajectories which form the 
Arnold web \cite{zas-book} which allows the Arnold diffusion 
to occur in higher-dimensional systems. This can be better 
understood when the FTLEs are plotted in the phase space 
projection $p_0^{(1)}\times p_0^{(2)}$, shown in Fig.~\ref{phase} 
for $K=0.15$ [see dashed line in Fig.\ref{coupled}(a)].
 \begin{figure}[htb]
 \unitlength 1mm
 \begin{center}
\includegraphics*[width=9.0cm,angle=0]{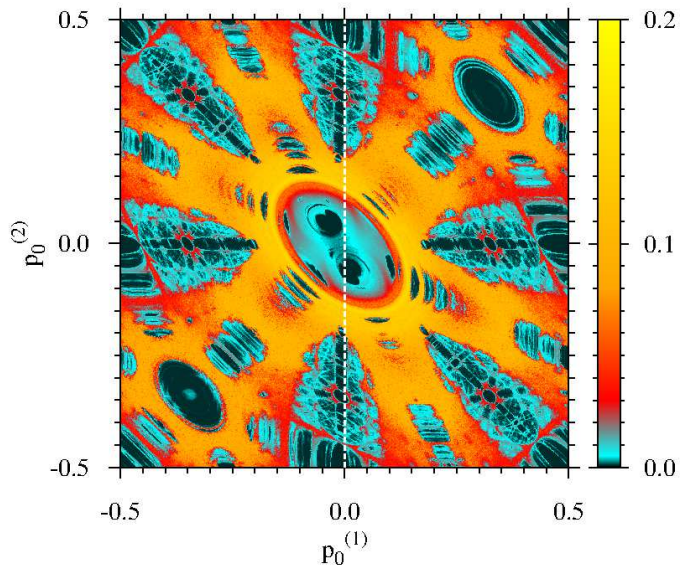}
 \end{center}
\caption{(Color online) FTLEs in the phase space projection
$p_0^{(1)}\times p_0^{(2)}$ for $K=0.15$ [see white dashed line
in Fig.~\ref{coupled}(a)].} 
  \label{phase}
  \end{figure}
Black points are related to close to zero FTLEs, while purple, red 
to yellow points are related to increasing FTLEs. This picture is 
similar to the one shown in \cite{kaneko94} (but in black and white), 
and presents a form called as ``onion''. In fact, this plot shows an 
Arnold web, or stochastic web, where points with larger FTLEs are 
the chaotic channels inside which the Arnold diffusion occurs. 
The FTLEs along the line $p_0^{(1)}=0.0$ in Fig.~\ref{phase} are 
exactly the FTLEs along the white dashed line in 
Fig.~\ref{coupled}(a). In other words, ICs for
$K=0.15$ [white dashed line in Fig.~\ref{coupled}(a)], inside 
the chaotic stripes, are those which generate trajectories with 
positive FTLEs and correspond exactly to the positive FTLEs from 
Fig.~\ref{phase} for $p_0^{(1)}=0.0$ (see also white dashed line).
Thus the chaotic stripes define the width of the chaotic channels,
for a given $K$, inside which Arnold diffusion occurs as time 
increases. Summarizing, the chaotic stripes in Fig.~\ref{coupled} 
show directly the intervals of ICs where the chaotic 
channels are located and how the Arnold web changes with $K$.


\section{Conclusions}
\label{conclusions}

Chaotic stripes are shown to generate weakly 
chaotic dynamics} for two distinct Hamiltonians systems, 
namely an open rectangular billiard with rounded corners 
and $4$ globally coupled particles on a discrete lattice. 
The width of the stripes increase with the nonlinear 
parameter and allow us to characterize the different domains 
of the dynamics. For the billiard case the stripes define 
injection angles which generate an chaotic dynamics inside
the billiard and an uncertainty about 
ejection angles.  For small values of the nonlinear 
parameter but higher-dimensional systems, the stripes 
are the chaotic channels in the Arnold web inside 
which the Arnold diffusion occurs \cite{zas-book}. 
From this perspective we can define the chaotic 
stripes as {\it Arnold stripes}. We also observed 
such stripes for Hamiltonian systems with higher 
values of $N>4$ and other couplings (not shown here).
This suggests that weak chaos in Hamiltonian systems 
always emerges inside stripes of particular 
ICs. This is of relevance since in 
such systems finite time distributions, independently
of the considered physical quantity, are not Gaussians
anymore and decay rates of recurrences and time 
correlations are difficult to obtain 
\cite{manchein09,cesarPRE09}. For higher 
values of the nonlinear parameters a totally chaotic 
motion is observed due to the overlap of Arnold stripes.

\begin{acknowledgments}
The authors thank CNPq, CAPES and FINEP (under project 
CT-INFRA/UFPR) for partial financial support.
\end{acknowledgments}


\end{document}